# On Decompositions of Lorentz Transformation and their physical interpretations.

by Chandru Iyer, Techink Industries C-42 Phase II, Noida, India – 201 305
e-mail: chandru.iyer@luxoroffice.com

Abstract:
In a recent article [1] we have explored alternative decompositions of the Lorentz transformation by adopting the synchronization convention of the target frame at the end and alternately at the outset. In this note we develop the decomposition by assuming a correct universal synchronization that may be outside the two inertial frames that are involved.

1. Introduction[1]: The measurement of the length of a co-moving object is unaffected by the synchronicity convention of an inertial frame. It is measured by an observer "by marking off his measuring-rod in a straight line as many times as is necessary to take him from one marked point to the other. Then the number which tells us how often the rod has to be laid down is the required distance" [2]. However, when an inertial frame measures the length of moving objects, it requires a set of spatially separated and synchronized clocks as it needs to determine the location of the two ends of the moving object at any particular instant. Hence the observed length of a moving object depends severely on the synchronicity convention adopted by an inertial frame. If we presume that the moving object does have a definite length, then there must be an associated synchronicity convention that correctly measures this length.

2. Various forms of the decomposition of Lorentz transformation

Let us suppose M and N are two inertial frames that are in relative motion and let us say that N is moving at v with respect to M as observed by M and M is moving at –v with respect to N as observed by N.

Let us assume that the internal synchronization of M and N are specific conventions of M and N. The correct universal synchronization is associated with another reference frame K. let us say M is moving at u with respect to K and N is moving at w with respect to K. It is to be noted that u and w are not observable either from M or N; however we presume that u and w have definite values even if unknown to us such that $v = \dfrac{w-u}{1-uw/c^2}$.

---

[1] This note is developed as a follow up to the previous article [1]; the introduction, references and other context is to be augmented from the same.

The correct decomposition of the transformation of event coordinates from M to N is given by M→K followed by K→N. As discussed in [1], the transformation M→K will take the form

$$\begin{pmatrix} \gamma_u & u\gamma_u \\ u\gamma_u/c^2 & \gamma_u \end{pmatrix} = \begin{pmatrix} 1 & u \\ 0 & 1 \end{pmatrix} \begin{pmatrix} 1/\gamma_u & 0 \\ 0 & \gamma_u \end{pmatrix} \begin{pmatrix} 1 & 0 \\ u/c^2 & 1 \end{pmatrix} \quad \ldots\ldots\ldots\ldots (1)$$

[Here we shift the synchronization first as we need to go to the correct universal synchronization of K at the outset]

and the decomposition from K to N will take the form.

$$\begin{pmatrix} \gamma_w & -w\gamma_w \\ -w\gamma_w/c^2 & \gamma_w \end{pmatrix} = \begin{pmatrix} 1 & 0 \\ -w/c^2 & 1 \end{pmatrix} \begin{pmatrix} \gamma_w & 0 \\ 0 & 1/\gamma_w \end{pmatrix} \begin{pmatrix} 1 & -w \\ 0 & 1 \end{pmatrix} \quad \ldots\ldots(2)$$

[Here we shift to the synchronization convention of N as the last process].

The overall decomposition will take the following form.

$$\begin{pmatrix} \gamma_v & -v\gamma_v \\ -v\gamma_v/c^2 & \gamma_v \end{pmatrix} = \left[\begin{pmatrix} 1 & 0 \\ -w/c^2 & 1 \end{pmatrix}\begin{pmatrix} \gamma_w & 0 \\ 0 & 1/\gamma_w \end{pmatrix}\begin{pmatrix} 1 & -w \\ 0 & 1 \end{pmatrix}\right] \left[\begin{pmatrix} 1 & u \\ 0 & 1 \end{pmatrix}\begin{pmatrix} 1/\gamma_u & 0 \\ 0 & \gamma_u \end{pmatrix}\begin{pmatrix} 1 & 0 \\ u/c^2 & 1 \end{pmatrix}\right]$$

----------------- (3a)

$$= \begin{pmatrix} \gamma_w & -w\gamma_w \\ -w\gamma_w/c^2 & \gamma_w \end{pmatrix} \begin{pmatrix} \gamma_u & u\gamma_u \\ u\gamma_u/c^2 & \gamma_u \end{pmatrix} \quad \ldots\ldots (3b)$$

Where $v = \dfrac{w-u}{1-uw/c^2}$ ……………………(4)

Equation (3a) can also be expressed as

$$\begin{pmatrix} \gamma_v & -v\gamma_v \\ -v\gamma_v/c^2 & \gamma_v \end{pmatrix} = \begin{pmatrix} 1 & 0 \\ -w/c^2 & 1 \end{pmatrix} \begin{pmatrix} \gamma_w/\gamma_u & (u-w)\gamma_w\gamma_u \\ 0 & \gamma_u/\gamma_w \end{pmatrix} \begin{pmatrix} 1 & 0 \\ u/c^2 & 1 \end{pmatrix} \quad \ldots\ldots\ldots (5a)$$

The inverse of above is given by,

$$\begin{pmatrix} \gamma_v & v\gamma_v \\ v\gamma_v/c^2 & \gamma_v \end{pmatrix} = \begin{pmatrix} 1 & 0 \\ -u/c^2 & 1 \end{pmatrix} \begin{pmatrix} \gamma_u/\gamma_w & (w-u)\gamma_w\gamma_u \\ 0 & \gamma_w/\gamma_u \end{pmatrix} \begin{pmatrix} 1 & 0 \\ w/c^2 & 1 \end{pmatrix} \quad \ldots\ldots\ldots (5b)$$

The above transformations are independent of the specific values of u and w as long as the relationship in equation (4) is valid.

We observe from equation (5a) and (5b) that the forward transformation (5a) has a linear deformation of ($\gamma_w/\gamma_u$) and the inverse has a linear deformation of ($\gamma_u/\gamma_w$). Thus they are physically consistent. Also the rate at which clocks run is altered by the factor ($\gamma_u/\gamma_w$) in the forward transformation and by the factor ($\gamma_w/\gamma_u$) in the inverse transformation. This is also physically consistent.

Further when $u = -w$ and $v = \dfrac{2w}{1+w^2/c^2}$

We have the transformation to be (from equation 5a)

$$\begin{pmatrix} \gamma_v & -v\gamma_v \\ -v\gamma_v/c^2 & \gamma_v \end{pmatrix} = \begin{pmatrix} 1 & 0 \\ -w/c^2 & 1 \end{pmatrix} \begin{pmatrix} 1 & -2w\gamma_w^2 \\ 0 & 1 \end{pmatrix} \begin{pmatrix} 1 & 0 \\ -w/c^2 & 1 \end{pmatrix} \quad \ldots\ldots\ldots (6)$$

[no contraction/elongation].

Where $v = \dfrac{2w}{1+w^2/c^2}$ and $w = (c^2/v)\left[1 - \sqrt{1-(v^2/c^2)}\right]$

The same transformation on the lhs of above equation can also be decomposed as below, which is the traditional representation.

$$\begin{pmatrix} \gamma_v & -v\gamma_v \\ -v\gamma_v/c^2 & \gamma_v \end{pmatrix} = \begin{pmatrix} 1 & 0 \\ -v/c^2 & 1 \end{pmatrix} \begin{pmatrix} \gamma_v & 0 \\ 0 & 1/\gamma_v \end{pmatrix} \begin{pmatrix} 1 & -v \\ 0 & 1 \end{pmatrix} \quad \ldots\ldots\ldots (7)$$

In equation [6] we see that the Lorentz transformation has a decomposition that does not involve any deformation of the rulers or slow/fast running of clocks. There is a shift in synchronization followed by Galilean motion and another shift in synchronization. This is actually the case where the universal synchronization is in between the two internal synchronizations of M and N. The rulers of both M and N are contracted w.r. to K and there is no relative contraction between M and N. Similarly the clocks of M and N are running at the same rate but slower than those of K. Here N is moving at +w, w.r.to K and N is moving at –w, w.r.to K the term $2w\gamma_w^2$ is the relative motion between M and N as observed by N and M with their contracted rulers and slow running clocks; it is 2w as observed by K. Whereas in the decomposition as given in equation (6), M and N do not observe any contraction of each others rulers, in the decomposition as given in equation (7), the rulers of M appear contracted as observed by N.

**3.** Conclusion: The observed length of a moving object depends severely on the synchronicity convention adopted by an inertial frame. If we presume that the moving object does have a definite length[2], then there must be an associated synchronicity convention that correctly measures this length. The fact that one has to make a shift in clock synchronization while transforming the event coordinates and that the length of a moving object depends on the chosen synchronization implies that one cannot be certain that the assumed internal synchronization of any frame is "correct". The decompositions given in equation (5a and 5b) above basically states that when we transform the event coordinates from frame M to N, we need to do a shift in synchronization that takes us to a universally correct synchronization, then a synchronized transformation that accounts for i)-movement, ii) deformation of liner dimensions of objects and iii) change in the rate of running of clocks and then shift to the E-synchronization of the target frame. In the special case when the universally correct synchronization lies in between M and N, there is no contraction or elongation of the rods of M and N as observed by each other using the universally correct synchronization. The clocks of M and N run at the same rate also as observed by each other using the universally correct synchronization. This is illustrated by the diagonal elements being unity in the middle matrix on the rhs of equation (6), namely $\begin{pmatrix} 1 & -2w\gamma_w^2 \\ 0 & 1 \end{pmatrix}$. However, when M and N observe each other using their own E-synchronization, they observe that the rods of each other are contracted and the clocks of each other run at a slower rate as given in equation (7).

In the general case also, as described in equations (5a) and (5b) and as already discussed in section 2, the forward and inverse transformations have a physically consistent manifestation of linear deformation and change in rate of running of clocks; that is these changes are reciprocal to each other in the forward and inverse transformations.

References:

1. Eur. J. Phys. 29 (2008) L13-L17.

---

[2] We presume that the length of a moving object can be observed to be a real number multiplied by the proper unit length defined in the observing frame.

2. Einstein, A. Relativity, The Special and General Theory, Authorized Translation by Robert W. Lawson, Three Rivers Press, New York pp32

Explanatory Notes are given in the next page in the form of a table.

# Explanatory Notes

| Sl. No. | Decomposition as given in | Physical interpretation of the decomposition |
|---|---|---|
| 1. | Equation (2) | This is the traditional decomposition where first movement is accounted for. Then there is a contraction of rulers and slow running of clocks. Then there is a shift to the E-synchronization of the target frame. |
| 2. | Equation (1) | In this case, the E-synchronization of the target frame is the correct synchronization. We first shift to the E-synchronization of the target frame; then there is elongation of rulers and faster running of clocks; then relative motion is accounted for. |
| 3. | Equation (3a) | In this case, the Lorentz transformation is first a synchronization shift from the reference frame to the "rest frame" accompanied by associated elongation and faster rate of running of clocks and then from the "rest frame" to the target frame, adopting the synchronization convention of the target frame at the very end. Thus it is a combination of rhs of equation (1) followed by rhs of equation(2). With u=0, this reduces to equation(2) and with w= 0, (whence v=-u), this reduces to equation(1). However, u and w can have any values such that $v = (w-u)/(1-(uw/c^2))$. This can be considered as the most general decomposition of the Lorentz transformation |
| 4. | Equation (3b) | The Lorentz transformation is propagative. That is $L_u + L_w = L_v$, where $v = u \oplus w$ |
| 5. | Equation (5a, 5b) | There is a universal synchronization to which we need to shift at the outset. Then we account for motion, deformation and rate of running of clocks in one single transformation. In the end we shift to E-synchronization of the target frame. This is an alternate form of equation 3(a). The linear deformation and change in rate of running of clocks in the inverse transformation are reciprocal to those in the forward transformation; thus these are physically consistent in both respects. |
| 6. | Equation (6) | This is a special case. Here the universal synchronization corresponding to the rest frame happens to be exactly in between the reference frame and the target frame. We shift to the universal synchronization first. Then there is only relative Galilean motion. Then we shift to the e-synchronization of the target frame. There is no relative contraction or elongation of the linear objects. Neither there is any change in the rate of running of clocks. However, linear objects of both the reference frame and target frame are contracted with respect to the rest frame. |